\documentclass[epj,nopacs,final]{svjour}
\usepackage{graphics}
\usepackage{latexsym,overpic,rotating,float}
\usepackage{subfigure,color,showkeys}
\usepackage{epsfig,color,rotating,amsmath,delarray,array,multirow}
\usepackage{makeidx,pifont,float,amssymb}
\usepackage{ulem} 
\newcommand{\er}{$\pm$}
\newcommand{\be}{\begin{eqnarray}}
\newcommand{\ee}{\end{eqnarray}}
\newcommand{\bea}{\begin{eqnarray}}
\newcommand{\eea}{\end{eqnarray}}
\newcommand{\bc}{\begin{center}}
\newcommand{\ec}{\end{center}}

\newcommand{\beq}{\begin{equation}}
\newcommand{\eeq}{\end{equation}}
\newcommand{\ba}{\begin{eqnarray}}
\newcommand{\ea}{\nonumber \end{eqnarray}}

\newcommand{\bi}{\begin{enumerate}}
\newcommand{\ei}{\end{enumerate}}
\def\MSL (#1,#2,#3,#4){[\Lambda\frac{#1}{2}^#2]_{#3}(#4)}
\def\MSS (#1,#2,#3,#4){[\Sigma\frac{#1}{2}^#2]_{#3}(#4)}

\begin{document}

\title{\boldmath $\Lambda$ and $\Sigma$ Excitations and the Quark Model}
\titlerunning{ $\Lambda$ and $\Sigma$ excitations and the Quark Model}

\author{E.~Klempt\inst{1,2}, V. Burkert\inst{2}, U. Thoma\inst{1}, L. Tiator\inst{3}, and  R. Workman\inst{4},\\[0.3ex]
The Baryon@PDG Group\\}
\authorrunning{E. Klempt {\it et al.}}

\institute{\inst{1}Helmholtz--Institut f\"ur Strahlen--
                     und Kernphysik, Universit\"at Bonn, 53115 Bonn, Germany\\
\inst{2}Thomas Jefferson National Accelerator Facility, Newport News, VA 23606\\
\inst{3}Institute of Nuclear Physics, Universit\"at Mainz, 55099 Mainz, Germany\\
\inst{4}Department of Physics, George Washington University, Washington, DC 20052, USA
}

\date{\today}

\abstract{The spectrum of $\Lambda$ and $\Sigma$ excitations is reviewed taking into account (nearly)
all hyperon resonances which were seen in early analyses or in one of the recent partial-wave analyses.
The spectrum is compared with the old Isgur-Karl model and the Bonn model. These models allows us
to discuss the SU(3) structure of the observed resonances. The SU(3) decomposition is compared with
SU(6) relations between the different decay modes. Seven $\Lambda$ states are proposed to be classified
as SU(3) singlet states. The hyperon spectrum is compared with the spectrum of $N$ and $\Delta$
resonances.}


\maketitle

\section{Introduction}
The spectrum of excited states of the nucleon and their internal structure are presently studied in
a number of laboratories, experimentally in photo- and electroproduction experiments, at the
phenomenological level in partial-wave analyses, and theoretically exploiting the quark model,
effective field theories or lattice gauge theory. SU(3) relates the spectrum of nucleon and
$\Delta$ resonances to the hyperon spectrum. In the $\Lambda$ spectrum, an additional class of
resonances turns up that are invariant under the exchange of all three quarks and that belong to
the SU(3) singlet. A comparison of the nucleon and hyperon excitation spectra should provide
information to what extent SU(3) symmetry holds and may help us in the spectroscopic interpretation
of resonances.

Most data on the hyperon spectrum were taken in bubble chambers studying $K^-$  induced reactions.
The hyperon spectrum based on early analyses can be found, e.g., in the 2012 Review of Particle Physics
 (RPP'2012)~\cite{Beringer:1900zz}. New data in the field of hyperon spectroscopy are scarce. There are
new data from  BNL covering the very low energy region of $K^-p$
scattering~\cite{Starostin:2001zz,Prakhov:2004ri,Prakhov:2004an,Manweiler:2008zz,Prakhov:2008dc}.
At Jefferson Lab (JLab) the spin and parity of the $\Lambda(1405)$ was determined \cite{Moriya:2014kpv} in a study
of the reactions $\gamma p \rightarrow K^+\Sigma^\pm \pi^\mp$ and $\gamma p \rightarrow K^+\Sigma^0
\pi^0$. The $\Lambda(1405)$  line shape was studied with real photons \cite{Moriya:2013eb}
and in electroproduction~\cite{Lu:2013nza}.
 A study of the production dynamics of low-mass hyperons was reported in
Ref.~\cite{Moriya:2013hwg}. Otherwise, no new data were reported since the 1980s. The status
remained at a stand-still for a long time.

A first break-through for our understanding of the baryon excitation spectrum was achieved in the
work of Isgur and Karl~\cite{Isgur:1978xj,Isgur:1978wd}. The Isgur-Karl model is based on a
non-relativistic Hamiltonian with a confinement potential for the constituent quarks and residual
quark-quark interactions via an effective one-gluon exchange; spin-orbit interactions were
suppressed. Its relativized version~\cite{Capstick:1986bm} returned similar results.
Other quark models followed: Glozman {\it et al.} considered the exchange of pseudoscalar mesons
between quarks (instead of one-gluon exchange)~\cite{Glozman:1995fu,Glozman:1997ag}. Hyperons were
classified into SU(3) flavor multiplets in Ref.~\cite{Melde:2008yr}. A relativistic quark-diquark
mass operator with direct and exchange interactions was suggested to solve the problem of the {\it
missing resonances}~\cite{Ferretti:2011zz,Santopinto:2014opa,Santopinto:2016fay}.  Coulomb-like interactions and a
confinement, both expressed in terms of a  hyperradius, were suggested to govern the dynamics of
quarks in baryons~\cite{Giannini:2015zia}. Faustov and Galkin calculated the hyperon mass spectra
in a relativistic quark model \cite{Faustov:2015eba} in an approximation which assumes that the two
light quarks form a diquark.  The Bonn model~\cite{Loring:2001ky,Loring:2001kv,Loring:2001kx} is
relativistically covariant and based on the Bethe-Salpeter equation with instantaneous two- and
three-body forces. The Isgur-Karl and the Bonn model give an expansion of the wave functions into
SU(3)-multiplets. We will compare the experimental spectrum with these two models.

Using lattice QCD, the masses of excited baryons that can be formed from $u$, $d$ and $s$ quarks
have been calculated~\cite{Edwards:2011jj,Edwards:2012fx}. The pattern of states is very similar to
the one obtained in quark models. Quark masses were used that correspond to a minimal pion mass
just below 400\,MeV.

Effective field theories (EFTs), when applied to baryons with $s$ quarks, concentrated on the role of
low-lying resonances in the $\bar KN$ $S$-wave. Kaiser, Waas and Weise~\cite{Kaiser:1996js}
constructed an effective potential from a chiral Lagrangian, and a resonance emerged as quasi-bound
state in the $\bar KN$ and $\pi\Sigma$ coupled-channel system: the $\Lambda(1405)1/2^-$ resonance.
Oller and Meissner~\cite{Oller:2000fj} studied the $S$-wave $\bar K N$ interaction between the
SU(3) octet of pseudoscalar mesons and the SU(3) octet of stable baryons in a relativistic chiral
unitary approach with coupled-channels and found two isoscalar resonances below 1450\,MeV, at about
1380\,MeV and 1434\,MeV. The first wider state was interpreted as mainly a singlet, a second state at
1434\,MeV  as mainly an octet state. A further state at 1680\,MeV was also identified
as mainly an octet state~\cite{Jido:2003cb}. The results were
confirmed in a number of further studies. A survey of the literature and a discussion of the
different approaches can be found in Ref.~\cite{Cieply:2016jby,Anisovich:2020tbd,Meissner:2020khl}.

In spite of the rareness of new data, new analyses have shed new light on the hyperon spectrum. The
Kent group (KSU) collected a large set of data on $K^-p$ interactions at low energies. The
partial-wave amplitudes were extracted~\cite{Zhang:2013cua} and fitted using a multichannel
parametrization consistent with S-matrix unitarity~\cite{Zhang:2013sva}. The KSU partial-wave
amplitudes were also fitted by the JPAC group in a coupled-channel fit \cite{Fernandez-Ramirez:2015tfa}.
The JPAC analysis was based on the K-matrix formalism, with special attention was paid to the analytical
properties of the amplitudes and their continuation to the complex angular momentum plane. The
Osaka-ANL group applied a dynamical coupled-channel approach, determined the resonances to achieve a
good fit and determined their properties \cite{Kamano:2014zba,Kamano:2015hxa}. Recently, the
Bonn-Gat\-china (BnGa) group increased the data set by adding further (old) data and reported the
hyperon spectrum \cite{Matveev:2019igl} and the properties of resonances~\cite{Sarantsev:2019xxm}.

\section{The full spectrum}

First, we discuss changes that have been introduced for the next RPP. Up to the 2018 edition, a few
resonances were reported that were observed as {\it bumps} in production experiments without a
partial wave analysis. Their masses were often close to known states. In the $\Sigma$ sector, there are also
several {\it bumps} that are likely produced by known states. In RPP'2020, the  $\Sigma(1480)$,
$\Sigma(1560)$, $\Sigma(1620)$, $\Sigma(1670)$, $\Sigma(1690)$ {\it bumps} are removed from
the Tables. Readers interested in these observations are referred to earlier RPP editions. In the
$\Lambda$ sector, further {\it bumps} are reported for high masses.
These states are claims for new states (even though they are without spin-parity determination)
in a mass range not covered by recent partial-wave analyses; they are kept in the Tables.

\begin{table}[pt]
\caption{\label{list-of-res}$\Lambda$ and $\Sigma$ resonances seen in early
analyses~\cite{Beringer:1900zz}, by the KSU~\cite{Zhang:2013sva}, the
Osaka-ANL~\cite{Kamano:2015hxa} collaboration,  the
JPAC~\cite{Fernandez-Ramirez:2015tfa}, and by the BnGa
collaboration~\cite{Sarantsev:2019xxm}.\vspace{-4mm}
} 
\renewcommand{\arraystretch}{1.3}
\bc
\begin{tabular}{llll}
\hline\hline
\hspace{-2mm}$\Lambda(1380)$ $1/2^-$&&
\hspace{-2mm}$\Sigma(1580)$ $3/2^-$&\hspace{-3mm}\cite{Beringer:1900zz,Fernandez-Ramirez:2015tfa}\cr
\hspace{-2mm}$\Lambda(1405)$ $1/2^-$&\hspace{-3mm}\cite{Beringer:1900zz,Zhang:2013sva,Sarantsev:2019xxm}&
\hspace{-2mm}$\Sigma(1620)$ $1/2^-$&\hspace{-3mm}\cite{Beringer:1900zz,Zhang:2013sva,Kamano:2015hxa,Fernandez-Ramirez:2015tfa,Sarantsev:2019xxm}\cr
\hspace{-2mm}$\Lambda(1520)$ $3/2^-$&\hspace{-3mm}\cite{Beringer:1900zz,Zhang:2013sva,Kamano:2015hxa,Fernandez-Ramirez:2015tfa,Sarantsev:2019xxm}&
\hspace{-2mm}$\Sigma(1660)$ $1/2^+$&\hspace{-3mm}\cite{Beringer:1900zz,Zhang:2013sva,Kamano:2015hxa,Fernandez-Ramirez:2015tfa,Sarantsev:2019xxm}\cr
\hspace{-2mm}$\Lambda(1600)$ $1/2^+$&\hspace{-3mm}\cite{Beringer:1900zz,Zhang:2013sva,Kamano:2015hxa,Fernandez-Ramirez:2015tfa,Sarantsev:2019xxm}&
\hspace{-2mm}$\Sigma(1670)$ $3/2^-$&\hspace{-3mm}\cite{Beringer:1900zz,Zhang:2013sva,Kamano:2015hxa,Fernandez-Ramirez:2015tfa,Sarantsev:2019xxm}\cr
\hspace{-2mm}$\Lambda(1670)$ $1/2^-$&\hspace{-3mm}\cite{Beringer:1900zz,Zhang:2013sva,Kamano:2015hxa,Sarantsev:2019xxm}&
\hspace{-2mm}$\Sigma(1750)$ $1/2^-$&\hspace{-3mm}\cite{Beringer:1900zz,Zhang:2013sva,Kamano:2015hxa,Sarantsev:2019xxm} \cr
\hspace{-2mm}$\Lambda(1690)$ $3/2^-$&\hspace{-3mm}\cite{Beringer:1900zz,Zhang:2013sva,Kamano:2015hxa,Fernandez-Ramirez:2015tfa,Sarantsev:2019xxm}&
\hspace{-2mm}$\Sigma(1775)$ $5/2^-$&\hspace{-3mm}\cite{Beringer:1900zz,Zhang:2013sva,Kamano:2015hxa,Fernandez-Ramirez:2015tfa,Sarantsev:2019xxm}\cr
\hspace{-2mm}$\Lambda(1710)$ $1/2^+$&\hspace{-3mm}\cite{Zhang:2013sva}&
\hspace{-2mm}$\Sigma(1780)$ $3/2^+$&\hspace{-3mm}\cite{Beringer:1900zz,Zhang:2013sva}\cr
\hspace{-2mm}$\Lambda(1800)$ $1/2^-$&\hspace{-3mm}\cite{Beringer:1900zz,Zhang:2013sva,Sarantsev:2019xxm}&
\hspace{-2mm}$\Sigma(1880)$ $1/2^+$&\hspace{-3mm}\cite{Beringer:1900zz,Zhang:2013sva}\cr
\hspace{-2mm}$\Lambda(1810)$ $1/2^+$&\hspace{-3mm}\cite{Beringer:1900zz,Zhang:2013sva,Kamano:2015hxa,Sarantsev:2019xxm}&
\hspace{-2mm}$\Sigma(1900)$ $1/2^-$&\hspace{-3mm}\cite{Beringer:1900zz,Zhang:2013sva,Kamano:2015hxa,Sarantsev:2019xxm}\cr
\hspace{-2mm}$\Lambda(1820)$ $5/2^+$&\hspace{-3mm}\cite{Beringer:1900zz,Zhang:2013sva,Kamano:2015hxa,Fernandez-Ramirez:2015tfa,Sarantsev:2019xxm}&
\hspace{-2mm}$\Sigma(1910)$ $3/2^-$&\hspace{-3mm}\cite{Beringer:1900zz,Sarantsev:2019xxm}\cr
\hspace{-2mm}$\Lambda(1830)$ $5/2^-$&\hspace{-3mm}\cite{Beringer:1900zz,Zhang:2013sva,Kamano:2015hxa,Fernandez-Ramirez:2015tfa,Sarantsev:2019xxm}&
\hspace{-2mm}$\Sigma(1915)$ $5/2^+$&\hspace{-3mm}\cite{Beringer:1900zz,Zhang:2013sva,Kamano:2015hxa,Fernandez-Ramirez:2015tfa,Sarantsev:2019xxm}\cr
\hspace{-2mm}$\Lambda(1890)$ $3/2^+$&\hspace{-3mm}\cite{Beringer:1900zz,Zhang:2013sva,Kamano:2015hxa,Fernandez-Ramirez:2015tfa,Sarantsev:2019xxm}&
\hspace{-2mm}$\Sigma(1940)$ $3/2^+$&\hspace{-3mm}\cite{Zhang:2013sva,Fernandez-Ramirez:2015tfa}\cr
\hspace{-2mm}$\Lambda(2000)$ $1/2^-$&\hspace{-3mm}\cite{Beringer:1900zz,Zhang:2013sva}&
\hspace{-2mm}$\Sigma(2010)$ $3/2^-$&\hspace{-3mm}\cite{Sarantsev:2019xxm}\cr
\hspace{-2mm}$\Lambda(2050)$ $3/2^-$&\hspace{-3mm}\cite{Zhang:2013sva,Fernandez-Ramirez:2015tfa}&
\hspace{-2mm}$\Sigma(2030)$ $7/2^+$&\hspace{-3mm}\cite{Beringer:1900zz,Zhang:2013sva,Kamano:2015hxa,Fernandez-Ramirez:2015tfa,Sarantsev:2019xxm}\cr
\hspace{-2mm}$\Lambda(2070)$ $3/2^+$&\hspace{-3mm}\cite{Sarantsev:2019xxm}&
\hspace{-2mm}$\Sigma(2070)$ $5/2^+$&\hspace{-3mm}\cite{Beringer:1900zz,Fernandez-Ramirez:2015tfa}\cr
\hspace{-2mm}$\Lambda(2080)$ $5/2^-$&\hspace{-3mm}\cite{Kamano:2015hxa,Fernandez-Ramirez:2015tfa,Sarantsev:2019xxm}&
\hspace{-2mm}$\Sigma(2080)$ $3/2^+$&\hspace{-3mm}\cite{Beringer:1900zz,Fernandez-Ramirez:2015tfa}\cr
\hspace{-2mm}$\Lambda(2085)$ $7/2^+$&\hspace{-3mm}\cite{Beringer:1900zz,Zhang:2013sva,Kamano:2015hxa,Fernandez-Ramirez:2015tfa}&
\hspace{-2mm}$\Sigma(2100)$ $7/2^-$&\hspace{-3mm}\cite{Beringer:1900zz,Fernandez-Ramirez:2015tfa,Sarantsev:2019xxm}\cr
\hspace{-2mm}$\Lambda(2100)$ $7/2^-$&\hspace{-3mm}\cite{Beringer:1900zz,Zhang:2013sva,Fernandez-Ramirez:2015tfa,Sarantsev:2019xxm}&
\hspace{-2mm}$\Sigma(2110)$ $1/2^-$&\hspace{-3mm}\cite{Zhang:2013sva,Sarantsev:2019xxm}\cr
\hspace{-2mm}$\Lambda(2110)$ $5/2^+$&\hspace{-3mm}\cite{Beringer:1900zz,Zhang:2013sva,Fernandez-Ramirez:2015tfa,Sarantsev:2019xxm}&
\hspace{-2mm}$\Sigma(2230)$ $3/2^+$&\hspace{-3mm}\cite{Sarantsev:2019xxm}\cr
\hspace{-2mm}$\Lambda(2325)$ $3/2^-$&\hspace{-3mm}\cite{Beringer:1900zz}&\\
\hspace{-2mm}$\Lambda(2350)$ $9/2^+$&\hspace{-3mm}\cite{Beringer:1900zz}&
\cr
\hline\hline
\end{tabular}\vspace{-4mm}
\ec
\end{table}
\begin{figure*}[pt]
\begin{center}
\includegraphics[width=\textwidth]{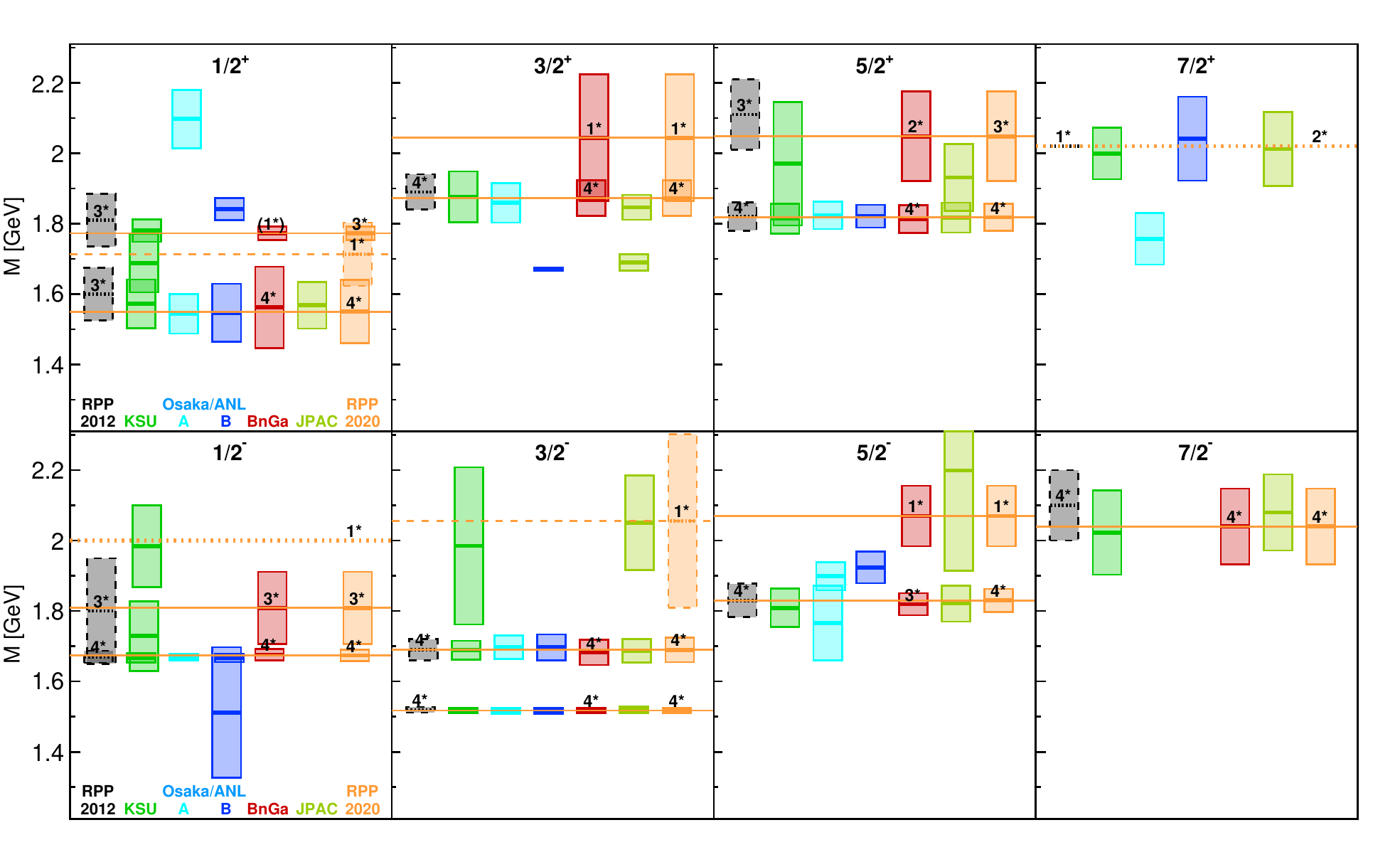}
 \vspace{-8mm}
\end{center}
\caption{\label{Lambdas} Mass spectrum of $\Lambda^*$ resonances above $\Lambda(1405)$ for
different spin and parities $J^P$. For each resonance, the real part of the pole position $Re(M_R)$
is given together with a box of length $\pm Im(M_R)$. $2\cdot Im(M_R)$ corresponds to the total
width of the resonance. The values of the KSU~\cite{Zhang:2013sva},
Osaka/ANL~\cite{Kamano:2015hxa}, BnGa~\cite{Matveev:2019igl,Sarantsev:2019xxm}, and
JPAC~\cite{Fernandez-Ramirez:2015tfa} analyses are given together with the values given in the
RPP'2012~\cite{Beringer:1900zz}
 and RPP'2020 listings. Star ratings are indicated for the RPP values and for the analyses, if
available. If no pole positions are given in the RPP (above the line), the RPP Breit-Wigner
estimates for masses and widths are used instead. This is indicated by dashed
resonance mass lines and dashed lines surrounding the boxes. If an RPP-
estimate is only available for the resonance mass but not for its width, no box
around the respective resonance mass is shown.
The RPP'2020 values are extended as lines throughout the picture to allow
for better comparison.
}
\end{figure*}

The more recent analyses
\cite{Zhang:2013sva,Kamano:2014zba,Kamano:2015hxa,Fernandez-Ramirez:2015tfa,Matveev:2019igl,Sarantsev:2019xxm} use, to a large
extent, the same data that had already been used in the early analyses reported in
Ref.~\cite{Beringer:1900zz}. However, the sets of extracted resonances are different. Which ones are
right? Consider, e.g., $\Sigma$ resonances in the 1850 to 1950\,MeV mass region. Early
analyses~\cite{Beringer:1900zz} find $\Sigma(1915)5/2^+$ and
$\Sigma(1910)3/2^-$$^($\footnote{This resonance was formerly called $\Sigma(1940)3/2^-$. It was renamed and is now called
$\Sigma(1910)3/2^-$ to avoid to have two
$\Sigma$ resonances with different $J^P$ but the same mass.}$^)$ as leading resonances
and additional evidence for $\Sigma(1880)$ $1/2^+$ and $\Sigma(1900)1/2^-$. 
The KSU group finds $\Sigma(1915)5/2^+$, $\Sigma(1900)1/2^-$, and $\Sigma(1940)3/2^-$, 
the Osaka-ANL group $\Sigma(1940)$ $1/2^-$,
$\Sigma(1890)5/2^+$, and BnGa finds $\Sigma(1915)$ $5/2^+$, $\Sigma(1900)1/2^-$, and $\Sigma(1910)3/2^-$.
All fits require the leading resonance $\Sigma(1915)5/2^+$, only the Osaka-ANL set does not include this state. But for the weaker resonances,
there are substantial differences. In the fits, resonances are usually added one by one. If a
resonance is added in this mass region, the fit improves when the quantum numbers are appropriate,
otherwise the fit does not improve, at least not significantly. However, when more and more
resonances are added, the improvement of the fit becomes smaller and smaller. When a good
description is reached, no further resonance is added. We assume, e.g., that if in the KSU
analysis, a $\Sigma(1910)3/2^-$ would have been tested, instead of $\Sigma(1940)3/2^+$, a gain in
fit quality would also have been reached. When it is added in addition to $\Sigma(1940)3/2^+$, the
quality of the fit improves only slightly. No significant improvement is obtained when a  $5/2^-$
resonance is  tested: a $5/2^-$ resonance does not exist in this mass region.
Thus both, $\Sigma(1910)3/2^-$ and $\Sigma(1940)3/2^+$, may exist but the
data are statistically not sufficient to reveal the existence of both at the same time. But
individually, they may both be uncovered. For these reasons, we consider all observations of a
resonance reported in one of the analyses in Refs.~\cite{Beringer:1900zz,Zhang:2013sva,Sarantsev:2019xxm},
or that were seen in model $A$ {\it and} $B$ in Ref. \cite{Kamano:2015hxa}, or that were considered a
trustworthy resonance in Ref.~\cite{Fernandez-Ramirez:2015tfa},
as candidates for a true state. In practice, this set mostly coincides with the RPP listings including all
resonances with at least one star.
In the KSU analysis, the partial-wave amplitudes are
constructed in sliced energy bins, and the resonance content in different partial waves seems to
be determined independently. However, the energy-indepen\-dent fit is guided by a first preliminary
energy-dependent fit. Table~\ref{list-of-res} summarizes the list of resonances.

\begin{figure*}[pt]
\begin{center}
\includegraphics[width=\textwidth]{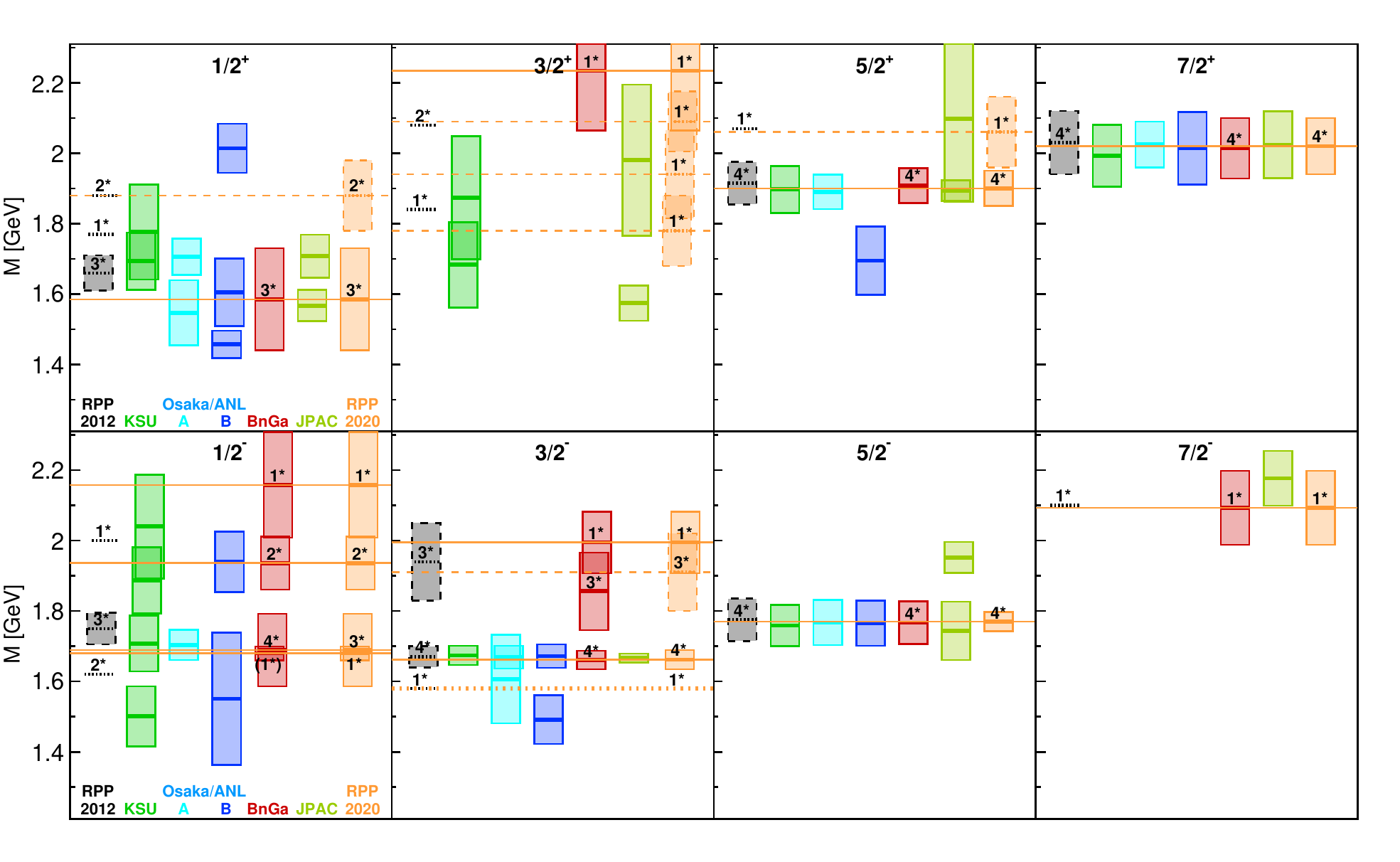}
 \vspace{-8mm}
\end{center}
\caption{\label{Sigmas}Mass spectrum of $\Sigma^*$ resonances above $\Sigma(1385)$. For further
explanations, see~Figure~\ref{Lambdas}.
}
\end{figure*}

The JPAC fit \cite{Fernandez-Ramirez:2015tfa} described the KSU partial waves~\cite{Zhang:2013sva}
reasonably well. However, when observables were calculated from their partial-wave amplitudes, significant
discrepancies \linebreak appeared. For this reason, the JPAC results were not included in the
RPP. Here, we include their spectrum of resonances in the discusssion (see Figs.~\ref{Lambdas}, \ref{Sigmas})
except those results that are marked by them
as unreliable or as artifacts of the fit.

Refs.~\cite{Beringer:1900zz,Zhang:2013sva,Fernandez-Ramirez:2015tfa,Sarantsev:2019xxm} assigned the
resonances found in their analyses to states listed in the RPP and gave unique values for masses
and widths. The Osaka-ANL group~\cite{Kamano:2015hxa} reported resonances found in their model $A$ or in their model $B$
but did not assign them to known states. In Table~\ref{list-of-res}, the Osaka-ANL observations are
associated with a known state or listed with a newly proposed name. Some resonances are seen in
both Osaka-ANL models, others only in their model $A$ or $B$. As Particle Data Group we decided to
list in the RPP only resonances which are seen in both model $A$ and $B$.

Fig.~\ref{Lambdas} shows the spectrum of all $\Lambda$ resonances observed in the different analyses,
here the results of model $A$ and $B$ from Ref.~\cite{Kamano:2015hxa} are shown.
The blocks give the pole masses and the pole widths of resonances. The horizontal lines indicate the
resonances and their masses adopted as a final spectrum.

The $\Lambda(1380)1/2^-$ and $\Lambda(1405)1/2^-$ resonances are below the $K^-p$ threshold and are
not reported in Refs.~\cite{Zhang:2013sva,Kamano:2015hxa,Fernandez-Ramirez:2015tfa,Sarantsev:2019xxm}.
When both
states exist, only one of them can be interpreted within the quark model. In this paper, we consider
$\Lambda(1405)1/2^-$ as the mainly SU(3) singlet state
and $\Lambda(1380)1/2^-$ as an intruder.

There are some cases where all measurements agree within a small band. This holds particularly true
for the well-known $\Lambda(1520)3/2^-$ but also for $\Lambda(1670)1/2^-$, $\Lambda(1690)$ $3/2^-$,
and $\Lambda(1815)5/2^+$ that all fall into a narrow mass band. Also very convincing are the
observations of  $\Lambda(1890)$ $3/2^+$, $\Lambda(1830)5/2^-$,
$\Lambda(2100)7/2^-$. These resonances are listed with four stars. The results on $\Lambda(1600)1/2^+$ are also rather
consistent; a four-star rating seems appropriate here as well.

%

$\Lambda(1800)1/2^-$, $\Lambda(1810)1/2^+$, and $\Lambda(2110)5/2^+$ are seen in several
early studies, and were listed as three-star resonances in~\cite{Beringer:1900zz}). They were
confirmed in the KSU, BnGa and partly in the Osaka-ANL (model B) analysis. These resonances
continue to be listed with three stars.
%

$\Lambda(2350)9/2^+$ is reported with three stars from early analyses. The recent analyses do not
cover this mass range.  Therefore, this resonance retains its status.
%

Under $\Lambda(2000)$, the RPP'2012~\cite{Beringer:1900zz} listed resonance claims near this
mass. Two entries under $\Lambda(2000)$ were reported with $J^P=3/2^-$ quantum
numbers~\cite{BarbaroGaltieri:1970xv,Brandstetter:1972xp} that were considered to be
obsolete~\cite{Beringer:1900zz}. They could have been shifted to
$\Lambda(2050)3/2^-$. The quantum numbers $J^P=1/2^-$ stem from
Refs.~\cite{Zhang:2013sva,Cameron:1978qi}.
%

The $\Lambda(2085)7/2^+$ was  seen in ~\cite{Beringer:1900zz}, and in the KSU, Osaka-ANL, and JPAC
analyses. The first mass determination of $\Lambda(2085)7/2^+$ was 2020
MeV~\cite{BarbaroGaltieri:1970xv} and that mass was used to give the resonance its name. The four
later determinations in the RPP found higher mas\-ses. The (unweighed) mean value of the five values
of Breit-Wigner masses is 2085\,MeV; therefore, in this study, we rename this state as
$\Lambda(2085)7/2^+$.

A $\Lambda(2080)5/2^-$ was seen in the Osaka/ANL analysis (model $A$ and $B$), by JPAC,
and by the BnGa group. The Osaka/ANL mass was considerably lower, the JPAC considerably
higher in mass compared to the BnGa result. We combined these observations to a single
one-star resonance.

The KSU and BnGa reported the states $\Lambda(1710)1/2^+$ and $\Lambda(2070)$ $3/2^+$, respectively. These
resonances is are not confirmed in other analyses and are listed as one-star resonances.

The Osaka/ANL group reports a $J^P=1/2^-$ $\Lambda^*$ resonance at 1512\,MeV. It is rather
wide, the pole width is 370\,MeV. We guess this pole might (miss-)represent the $\Lambda(1405)1/2^-$.

%
%

Some $\Sigma$ resonances like $\Sigma(1670)3/2^-$, $\Sigma(1775)5/2^-$, $\Sigma(1915)5/2^+$, and
$\Sigma(2030)7/2^+$ are seen with good consistency. They have a four-star status.
%

The $\Sigma(1660)1/2^+$ is consistently needed in all analyses. Its Breit-Wigner mass is
considerably larger than the pole mass; the pattern resembles the one of the Roper resonance that
has a large Breit-Wigner mass and a smaller pole mass. The spread of the values for its mass and width
are comparatively large, hence we keep its three-star status.
%
%

The $\Sigma(1620)1/2^-$ was given a two-star status in RPP 2012~\cite{Beringer:1900zz},
the $\Sigma(1750)1/2^-$ was listed with three stars. KSU finds two states. The low-mass pole
has a mass of 1501\,MeV and a (full) pole width of 171\,MeV, the high-mass pole has parameters 1708\,MeV and
158\,MeV, respectively. Osaka/ANL finds either a low-mass pole (model A) with a substantial width,
or (model B) a high-mass pole at 1764$^{+3}_{-6}$\,MeV and a full width of 84$^{+14}_{- 4}$\,MeV
(model A). BnGa finds a problematic - and statistically not significant -
 interference pattern of $\Sigma(1620)1/2^-$ and $\Sigma(1750)1/2^-$. Nevertheless, we keep the
three-star status $\Sigma(1750)1/2^-$ but decided to downgrade $\Sigma(1620)1/2^-$ to one star.
%

\ \ The $\Sigma(1880)1/2^+$ was given a two-star status in RPP'2012~\cite{Beringer:1900zz}. Several
other analyses do not confirm this state. Nevertheless, we keep it as two-star resonance.
%

We exclude the $\Sigma(1770)1/2^+$ resonance from the Listings. This resonance had been found in a
few early analyses. However, the positive result in Ref.~\cite{Kane:1972qa} was superseded by
Ref.~\cite{Kane:1974gg} where the resonance was found with a mass compatible with
$\Sigma(1660)1/2^+$. Its observation in Ref.~\cite{Gopal:1976gs} was superseded in
Ref.~\cite{Gopal:1980ur}, again a mass  compatible with $\Sigma(1660)1/2^+$ was found. The
resonance was also reported to exist in Ref.~\cite{Baillon:1975sk}, but only one of two solutions
required it.
%

In most cases, $\Lambda$ and $\Sigma$ resonances reported only in RPP'2012~\cite{Beringer:1900zz}
or only either in the KSU~\cite{Zhang:2013sva} or in the BnGa~\cite{Sarantsev:2019xxm} analysis are
listed as one-star resonances.
%

The one-star $\Sigma(1580)3/2^-$ was seen in an analysis of low-statistics data on
$K^-p\to\pi^0\Lambda$~\cite{Litchfield:1974ru} but has been ruled out by a counter experiment at
BNL~\cite{Prakhov:2004an,Olmsted:2003is}. It was confirmed in the Osaka/ANL analysis.
%

The $\Sigma(1840)3/2^+$ resonance was removed from the Listings. It contains results from early
analyses. Two entries are compatible with $\Sigma(1780)$ $3/2^+$
and one entry is compatible with
$\Sigma(1940)$ $3/2^+$. These entries are moved to the corresponding resonances.

The $\Sigma(1910)3/2^-$ is a three-star resonance long known as the $\Sigma(1940)3/2^-$. Recently, it was confirmed
in the BnGa analysis. It was renamed to avoid confusion with $\Sigma(1940)3/2^+$.


The $\Sigma(2000)1/2^-$ has been removed from the listings and the results are transferred to
$\Sigma(1900)1/2^-$, the mass value derived in Ref.~\cite{Zhang:2013sva}. The
resonance was confirmed in the BnGa analysis.
The entries are mostly compatible with mass values in the 1900 to 1950\,MeV range.
%

Resonances reported in none of the three
papers~\cite{Beringer:1900zz,Zhang:2013sva,Sarantsev:2019xxm} or only in Osaka-ANL-model $A$ or $B$
are not taken into account. This holds true, e.g., for $\Lambda(1687)3/2^+$ and $\Sigma(1574)3/2^+$
in the JPAC analysis, $\Lambda(1757)7/2^+$ in Osaka-ANL-model $A$,  and $\Lambda(2097)7/2^+$,
$\Lambda(1671)$ $3/2^+$, $\Sigma(1695)$ $5/2^+$ in Osaka-ANL-model $B$. Furthermore, these states are certainly
incompatible with any quark model.  They might provide hints for resonances beyond the quark model,
or they could be artifacts of the fit. We disregard them in the further evaluations.

%
%

\section{\label{SU6}Symmetry considerations}
In the quark model, the wave function  of a baryon contains three constituent quarks. The wave
function is the product of four parts describing the spatial, spin, flavor, and color
configuration. The Pauli principle requires the baryon to be antisymmetric with respect to the
exchange of any pair of quarks. Confinement requires the color wave function to be
antisymmetric, hence the spatial-spin-flavor wave function needs to be symmetric.

For baryons with light-quark flavors (\textit{i.e.} $up, down$, and $strange$ ($u,d,s$) quarks)
only, the baryon flavor wave function can be decomposed into a decuplet which is symmetric with
respect to the exchange of any two quarks, a singlet which is antisymmetric and two octets of mixed
symmetry. The latter can be further classified into a mixed symmetric and a mixed antisymmetric
representation:
\begin{equation}
  3 \otimes 3 \otimes 3 = 10_\mathcal{S} \oplus 8_\mathcal{M} \oplus 8_\mathcal{M} \oplus 1_\mathcal{A}.
\end{equation}
$\Lambda$ hyperons can be in the flavor SU(3) singlet or octet, $\Sigma$ hyperons in the SU(3)
octet or decuplet.

The spin wave function can be classified according to SU(2) representations, the spin-flavor wave
function can be classified according to ${\rm SU(2)\otimes SU(3)=SU(6)}$-represen\-tations:
\begin{equation}
  6 \otimes\ 6 \otimes\ 6
  =
  56_\mathcal{S} \oplus\ 70_\mathcal{M} \oplus\ 70_\mathcal{M} \oplus\ 20_\mathcal{A}\,.
\end{equation}
The states in each of these representations can be decomposed
according to ${\rm SU(2)\otimes SU(3)}$,
\begin{equation}
  56
  =
  {^4}10 \oplus {^2}8.
  \label{56}
\end{equation}
\textit{i.e.} into a flavor symmetric decuplet combined with a spin symmetric quartet and the
symmetric combination of mixed symmetric flavor octet and mixed symmetric spin doublet states. This
spin-flavor wave function can be combined with a symmetric spatial wave function.

\begin{table*}[pt]
\caption{\label{Bonn-Lam}Configuration mixing of $\Lambda$ resonances in the
Isgur-Karl~\cite{Isgur:1978xj,Isgur:1978wd} (first number) and in Bonn model
$\mathcal{A}$~\cite{Loring:2001ky} (second number); the fractions are given in \%, n.g. = not
given, * = our value (assuming no mixing with the third excitation band). 
Shown are negative-parity states in the $1\hbar\omega$ band and positive-parity states in
the $2\hbar\omega$ band. In the $2\hbar\omega$ band, six states with $J^P=1/2^+$ are expected,
seven with $J^P=3/2^+$, five with $J^P=5/2^+$, and one with $J^P=7/2^+$. The table includes only
the lower-mass states; higher-mass states are omitted. The dominant fraction is highlighted.}
\begin{center}
\renewcommand{\arraystretch}{1.6}
\begin{tabular}[h]{|c||c||cccc||cc||c|}
\hline
$J^\pi$&
Model state &
$\!\!{}^2 8[56]\!\!$&
$\!\!{}^2 8[70]\!\!$&
$\!\!{}^4 8[70]\!\!$&
$\!\!{}^2 8[20]\!\!$&
$\!\!{}^2 1[70]\!\!$&
$\!\!{}^4 1[20]\!\!$
& Identified with\\[1mm]
\hline
\hline
                      &$\MSL(1,-,1,1490/1524)$&n.g./2.9&18/26.0&0/0.3&n.g./0.0&{\bf\underline{81/69.4}}&n.g./0.2&$\Lambda(1405)1/2^-$\\
\cline{2-9}
$\frac{1}{2}^-$&$\MSL(1,-,2,1650/1630)$&n.g./5.5&{\bf\underline{55/61.6}}&34/2.1&n.g./0.3&15/29.2&n.g./0.1&$\Lambda(1670)1/2^-$\\
\cline{2-9}
                      &$\MSL(1,-,3,1800/1816)$&n.g./0.2&25/3.1&{\bf\underline{72/94.9}}& n.g./0.6&3/0.1&n.g./0.6&$\Lambda(1800)1/2^-$\\
\hline
\hline
                      &$\MSL(3,-,1,1490/1508)$&n.g./2.0 &16/18.7&0.0/0.1&n.g./0.0&{\bf \underline{83/77.7}}&n.g./0.1&$\Lambda(1520)3/2^-$\\
\cline{2-9}
$\frac{3}{2}^-$&$\MSL(3,-,2,1690/1662)$&n.g./4.4&{\bf \underline{83/72.0}}&1/2.2&n.g./0.2&16/20.1&n.g./0.0&$\Lambda(1690)3/2^-$\\
\cline{2-9}
                      &$\MSL(3,-,3,1880/1775)$&n.g./0.8&1/1.5&{\bf \underline{98/96.1}}&n.g./0.0&0/0.4&n.g./0.4&\\
\hline
\hline
$\frac{5}{2}^-$&$\MSL(5,-,1,1815/1828)$&0*/0.0&0*/0.0&{\bf \underline{100*/99.0}}&0*/0.0&0*/0.0&0*/0.4&$\Lambda(1830)5/2^-$\\
\hline
\hline
\cline{2-9}
                     &$\MSL(1,+,1,1555/1677)$&{\bf\underline{99/88.4}} &0/6.2&0/0.1&0/0.2&1/3.7&0/0.1&$\Lambda(1600)1/2^+$\\
\cline{2-9}
                     &$\MSL(1,+,2,1740/1747)$&1/5.1&9/2.1&0/0.0&0/0.1& {\bf \underline{90/90.6}}&0/0.9&$\Lambda(1710)1/2^+$\\
\cline{2-9}
$\frac{1}{2}^+$&$\MSL(1,+,3,1860/1898)$&0/9.1&{\bf \underline{83/84.2}}&8/1.0&0/0.8&8/3.8&0/0.2&$\Lambda(1810)1/2^+$\\
 \cline{2-9}
                     &$\MSL(1,+,4,2020/2077)$&0/0.5&8/1.2&{\bf\underline{72/85.8}}&19/11.2&1/0.1&1/0.3&\\
\hline
\hline
                      &$\MSL(3,+,1,1810/1823)$&{\bf \underline{62/60.0}}&24/28.2&1/0.3&0/0.1&12/9.9&1/0.1&$\Lambda(1890)3/2^+$\\
 \cline{2-9}
$\frac{3}{2}^+$&$\MSL(3,+,2,1960/1952)$&1/3.8&8/7.6&31/0.8&0/0.1&\bf \underline{47/84.0}&11/2.2&$\Lambda(2070)3/2^+$\\
 \cline{2-9}
                      &$\MSL(3,+,3,2005/2045)$&12/0.5&0/0.2&{\bf\underline{60/96.9}}&1/1.1&20/0.3&9/0.2&\\
\hline
\hline
                     &$\MSL(5,+,1,1815/1834)$&{\bf \underline{65/57.8}}&23/28.3&0/0.2&0/0.1&12/12.1&0/0.0&$\Lambda(1820)5/2^+$\\
\cline{2-9}
$\frac{5}{2}^+$&$\MSL(5,+,2,2010/1999)$&2/4.5&18/8.9 &0/1.0 &0/0.1&{\bf \underline{79/84.1}}&1/0.2&$\Lambda(2110)5/2^+$\\
\cline{2-9}
                     & $\MSL(5,+,3,2095,2078)$&28/9.0&{\bf\underline{48}}/9.9&15/{\bf\underline{77.1}}&0/0.0&8/2.0&1/0.9&\\
\hline
\hline
$\frac{7}{2}^+$&$\MSL(7,+,1,2070/2130)$&0/0.0&0/0.0&{\bf \underline{100/99.1}}&0/0.0&0/0.0&0/0.1&$ \Lambda(2085)7/2^+$\\
 \hline
\hline
\end{tabular}
\vspace{-4mm}
\end{center}
\end{table*}

The mixed symmetric 70-plet can be decomposed as
\begin{equation}
  70
  =
  {^2}10 \oplus {^4}8 \oplus {^2}8 \oplus {^2}1.
  \label{70}
\end{equation}
It needs to be combined with a spatial wave function of mixed symmetry. A ground state has a
symmetric spatial wave function, hence SU(3) singlet $\Lambda$ states always carry orbital or
radial excitation.

Finally, the antisymmetric 20-plet contains the antisymmetric combination of a flavor mixed
symmetric octet with a mixed symmetric spin doublet and the antisymmetric flavor singlet combined
with a symmetric spin quartet:
\begin{equation}
  20
  =
  {^2}8\ \oplus\ {^4}1.
  \label{20}
\end{equation}

Nucleon and $\Delta$ resonances are easily identified: $\Delta$ exist in four charge states
$\Delta^{-}$, $\Delta^{0}$, $\Delta^{+}$, $\Delta^{++}$;  nucleons only in two charge states. Their
decays are governed by the well-conserved isospin symmetry. $\Lambda$ resonances in SU(3) singlets
or octets are always neutral in charge, $\Sigma$ resonances in octets or decuplets are found in
three charge states. A classification according to their SU(3) structure suffers from two aspects:
$\Lambda$ resonances in SU(3) singlet or octet configurations -- or $\Sigma$ resonances in SU(3)
octet or decuplet configurations -- can mix; second, SU(3) symmetry is significantly broken.

\section{\label{Conf-mix-Lam}Configuration mixing of $\Lambda$ resonances}
\subsection{\label{CompLam}Comparison with quark models}

The resulting spectrum is now compared to the classical Isgur-Karl model
\cite{Isgur:1978xj,Isgur:1978wd} and the later Bonn model~\cite{Loring:2001ky}. In
Table~\ref{Bonn-Lam}, $\Lambda$ resonances in the first and second excitation band are listed in
each partial wave by assuming that all resonances have been seen up to a maximum mass. Not all
resonances in the second band are listed; entries with high mass values  for which no experimental
candidates exist are not included in Table~\ref{Bonn-Lam} (nor in Table~\ref{Bonn-Sig}). In
particular, resonances belonging to the 20plet are all missing. The configuration of resonances in
the third band and in higher bands is not given in the publications.

Some $\Lambda$ resonances with a given spin-parity $J^P$ may have contributions from
quark-spin-doublets or quark-spin-quar\-tets, they may have a symmetric or antisymmetric spatial
wave function or a wave function  with mixed symmetry. They can be in a flavor singlet or octet
state. These internal quantum numbers are not observable, hence all these configurations can mix
when they have the same spin-parity. Table~\ref{Bonn-Lam} gives the probability that a physical
state with defined $J^P$ has a given set of internal quantum numbers. Ref.~\cite{Loring:2001ky}
uses a fully relativisic treatment and a small fraction of the wave function is found at negative
energy; these fractions are omitted here. In Refs.~\cite{Isgur:1978xj,Isgur:1978wd}, amplitudes are
given from which we calculated the probabilities. In Ref.~\cite{Loring:2001ky}, the two configurations 
$^48[70]$ with intrinsic $S$-wave or $D$-wave orbital excitation are not given separately. To allow
for an easier comparison, the two contributions calculated in Ref.~\cite{Isgur:1978wd} are added.

Given the large differences in the model assumptions, the agreement between the two models is
remarkable. In particular, the largest contributions, underlined in Table~\ref{Bonn-Lam}, are
mostly the same in both calculations. In \linebreak the negative-parity sector, the doublet
$\Lambda(1405)1/2^-$ and $\Lambda(1520)3/2^-$ have a large fraction in the $^21[70]$ configuration,
$\Lambda(1670)1/2^-$ and $\Lambda(1690)3/2^-$ are dominantly $^28[70]$, and $\Lambda(1800)1/2^-$
and $\Lambda(1830)5/2^-$ belong to a spin-quartet $^48[70]$ (degenerated to a triplet) where the
$3/2^-$ state is missing. The octet $\Lambda$ resonances have all analogue states in the nucleon
sector that are about 100 to 150\,MeV lower in mass.

In the positive-parity sector, the  $\Lambda(1600)1/2^+$ plays the role of the Roper
$N(1440)1/2^+$, the  $\Lambda(1810)1/2^+$  the role of $N(1710)1/2^+$. Between these two states,
there is a further state, $\Lambda(1710)1/2^+$, that cannot be mapped onto nucleon spectrum and is
interpreted as SU(3) singlet state in both quark models.

The two resonances $\Lambda(1890)3/2^+$ and $\Lambda(1820)5/2^+$ can be assigned to states that
belong - with a $\approx60$\% probability - to a $^28$ configuration in the 56-plet; $N(1720)3/2^+$ and
$N(1680)$ $5/2^+$ are their partners in the nucleon sector.  

The $\Lambda(2085)$ $7/2^+$ is a bit
low in mass when compared to $N(1990)7/2^+$; however, their spectroscopic identification is uni\-que:
both resonances must belong to the $^48[70]$-plet.

A slight discrepancy between the Isgur-Karl and the Bonn model in the spectroscopic assignment is
found for the $\Lambda(2070)3/2^+$ resonance. In the Bonn model,
this state is predominantly a singlet state, in the $^21[70]$-plet. In the Isgur-Karl model,
it is strongly mixed even though the $^21[70]$-plet configuration prevails. The $\Lambda(2110)5/2^+$ 
seems to be the spin partner. The next-higher
state with $J^P=5/2^+$ is predicted at 2095 or 2078\,MeV, depending on the model, it could be in a 
spin doublet or spin-quartet.

\subsection{\label{sec:decays}SU(3) constraints for $\Lambda^*$ decays}

The decays of hyperons are governed by the available phase space, the angular-momentum barrier, and
by symmetric ($d_{ijk}$) and antisymmetric ($f_{ijk}$) SU(3) structure constants. These are
tabulated, e.g., in the RPP. Their relative contribution is governed by the so-called $F/D$ ratio
that is usually parameterized as $F/D=\alpha/(1-\alpha)$.  In SU(3), $\alpha$ is a free parameter
but within SU(6), $\alpha$ can be predicted. With the values for $\alpha$ given in
Table~\ref{tab:L-decays}, the SU(6) coupling constants can be calculated (see, e.g.,
\cite{Guzey:2005rx,Guzey:2005vz}). The corresponding coupling constants are listed in
Table~\ref{tab:L-decays}. The initial state cancels in the comparison, and the relative sign of the
amplitudes can be used to determine the SU(3) structure of a hyperon.

\begin{table}[pt]
\caption{\label{tab:L-decays}SU(3) coupling constants for hyperon decays and the SU(6) predictions
for the coefficient $\alpha$ in decays of octet hyperons.}
\renewcommand{\arraystretch}{1.3}
\bc
\begin{tabular}{cccc}
\hline\hline
Decay mode                &$8\to 8+8$ & $1\to 8+8$           & \\\hline
$\Lambda\to N\bar K$  &$\sqrt{\frac23}(2\alpha +1)A_8$& $\frac12 A_1$       & \\
$\Lambda\to \Sigma\pi$&$2(\alpha-1)A_8$             &$\sqrt{\frac32}A_1$& \\\hline
&$^28[56]$& $^28[70]$ & $^48[70]$ \\
$\alpha$&$\frac25$ & $\frac58$ & $-\frac{1}{2}$\\
\hline
\end{tabular}

\renewcommand{\arraystretch}{1.5}
\begin{tabular}{ccccc}
&$^21[70]$&$^28[56]$& $^28[70]$ & $^48[70]$ \\\hline
$\frac{A(\Lambda\to N\bar K)}{A(\Lambda\to \Sigma\pi)}$&$\sqrt{\frac16}$&  $-\sqrt{\frac32}$&   $-\sqrt6$ &    0   \\
Sign &$+$&$-$&$-$&\\
\hline\hline
\end{tabular}
\vspace{-4mm}
\ec
\end{table}

The ratios of the decay amplitude for decays of $\Lambda$ resonances into $N\bar K$ and $\Sigma\pi$,
derived from SU(3) relations and imposing constrainst from SU(6), cannot be expected to be
fulfilled. This can best been seen in heavy-quark baryons. In the $\Xi_b$, the $u$ and the
$s$-quark are antisymmetric with respect to their exchange, in the $\Xi_b'$, the $u$ and $b$-quark
are antisymmetric. The mass difference between the two states is 110\,MeV. A potential ''SU(3)
symmetry in the $u,s,b$-quark sector'' is heavily broken, a better basis is the flavor
$u,s,b$-quark basis. This limit is called ``ideal mixing for baryons''. Likewise, the SU(3)
symmetry is broken in the $u,d,s$-quark sector and possibly, the physical states do not respect a
symmetry in which $u,d,s$-quarks can arbitrarily be exchanged but are better described in a $u,d,s$
basis. Thus, we cannot expect branching ratios to respect SU(6) symmetry. Experimentally, however,
the relative sign of the amplitude often signals the singlet or octet status of a $\Lambda$
resonance. Hence we discuss here only the relative sign of the amplitudes for $K^-p\to\Lambda^*\to
N\bar K$ and $K^-p\to\Lambda^*\to\Sigma\pi$.

\begin{table}[pt]
\caption{\label{Phases-Lam} The phase difference $\delta\phi$ between the amplitudes for
$K^-p\to\Lambda^*\to\Sigma\pi$ and $K^-p\to\Lambda^*\to K^-p$ on the real axis (2012, KSU) and at
the $\Lambda^*$ pole (Osaka-ANL and BnGa). 2012 stands for results listed in
Ref.~\cite{Beringer:1900zz}; in the case of inconsistent results, the signs reported more often are 
given here. The
relative sign or phase is not given in Ref.~\cite{Fernandez-Ramirez:2015tfa}. For the Osaka-ANL
(OA), the results from model $A$ and $B$ are quoted. A $-$ sign indicates that the resonance was
not reported. Resonances belonging to a spin quartet are predicted not to couple to $\bar KN$ when
they do not mix with other configurations. See text for which phases are to be ``expected''. }
\bc
\renewcommand{\arraystretch}{1.25}
\begin{tabular}{cccccc}
\hline\hline
\multirow{2}{*}{$\Lambda^*$}      &\multicolumn{5}{c}{Expected: $\delta\phi=0^\circ$} \\
\cline{2-6}&2012&KSU&OA$_A$&OA$_B$&BnGa\phantom{zz}\\\hline
$\Lambda(1520)3/2^-$  &$-$&    0$^\circ$&  1$^\circ$ &1$^\circ$&   -(5\er4)$^\circ$ \\
$\Lambda(1710)1/2^+$ &$-$&180$^\circ$&$-$&    $-$             &   $-$  \\
$\Lambda(2070)3/2^+$ &$-$&  $-$          &$-$&    $-$            &-(10\er13)$^\circ$ \\
$\Lambda(2080)5/2^-$  &$-$&     $-$& -&   -              &(46\er22)$^\circ$ \\
$\Lambda(2100)7/2^-$  & 0$^\circ$&    0$^\circ\dagger$& $-$         &$-$&  (5\er18)$^\circ$ \\
$\Lambda(2110)5/2^+$ & 0$^\circ$&    0$^\circ$&$-$&   $-$             &  -(5\er21)$^\circ$ \\
\hline
\multirow{2}{*}{$\Lambda^*$}   &\multicolumn{5}{c}{Expected: $\delta\phi=\pm180^\circ$}\\
\cline{2-6}&2012&KSU&OA$_A$&OA$_B$&BnGa\phantom{zz}\\\hline
$\Lambda(1600)1/2^+$ & 180$^\circ$&    180$^\circ$& -172$^\circ$ &-139$^\circ$& -(149\er14)$^\circ$ \\
$\Lambda(1670)1/2^-$  & 180$^\circ$&    180$^\circ$&-39$^\circ$   & -29$^\circ$&-(70\er18)$^\circ$  \\
$\Lambda(1690)3/2^-$  & 180$^\circ$&    180$^\circ$& -176$^\circ$ &  -174$^\circ$&-(157\er8)$^\circ$  \\
$\Lambda(1810)1/2^+$ & 180$^\circ$&    180$^\circ$&   $-$           & 40$^\circ$&(152\er35)$^\circ$  \\
$\Lambda(1820)5/2^+$ & 180$^\circ$&    180$^\circ$&-179$^\circ$ &   -177$^\circ$&-(164\er7)$^\circ$  \\
$\Lambda(1890)3/2^+$ & 180$^\circ$&   180$^\circ$& 127$^\circ$  & $-$ &(148\er16)$^\circ$  \\
$\Lambda(2000)1/2^-$  &180$^\circ$& 180$^\circ$  &   $-$              &$-$& $-$ \\
$\Lambda(2050)3/2^-$ & 180$^\circ$&0$^\circ$&   $-$             &$-$& $-$ \\
\hline
\multirow{2}{*}{$\Lambda^*$}   &\multicolumn{5}{c}{undefined}\\
\cline{2-6}&2012&KSU&OA$_A$&OA$_B$&BnGa\phantom{zz}\\\hline
$\Lambda(1800)1/2^-$  & 180$^\circ$&    180$^\circ$&   $-$             &$-$&(134\er11)$^\circ$  \\
$\Lambda(1830)5/2^-$  & 180$^\circ$&    180$^\circ$&-101$^\circ$&$2^\circ$&(160\er17)$^\circ$  \\
$\Lambda(2085)7/2^+$ & 180$^\circ$&0$^\circ$&  $-$            &2$^\circ$  & $-$ \\
 \hline\hline
\end{tabular}
\vspace{-4mm}
\ec
\end{table}

Table~\ref{Phases-Lam} lists the phases given in Refs.~\cite{Zhang:2013sva,Kamano:2015hxa}
and~\cite{Sarantsev:2019xxm}. The results  are organized into three blocks, one in which phases are
given for resonances that are -- based on Table~\ref{Bonn-Lam} -- supposed to be SU(3) singlets,
the second one contains resonances supposed to be in SU(3) octets. The third block lists octet resonances
assigned to the $^4[70]$-plet where the SU(6) phase is not defined.  Rescattering in the final state
and background contributions can lead to a shift of the phase; we interpret values which are
compatible the range $-35^\circ<\delta<35^\circ$ as  consistent with $\delta=0^\circ$ and the range
$145^\circ<\delta<215^\circ$ as consistent with $\delta=180^\circ$.

The singlet candidates have mostly phases which are compatible with this assignment. The phase
difference of the amplitude for $K^-p\to \Lambda(1520)3/2^- \to \bar KN$ and $\to \Sigma\pi$ is
$\delta=0^\circ$ in the KSU analysis. Osaka-ANL and BnGa determined the phase difference of the
normalized residues at the pole position and found $\delta=1^\circ$ or $\delta= (5\pm4)^\circ$,
respectively, compatible with $0^\circ$. Hence $\Lambda(1520)3/2^-$ is dominantly a SU(3) singlet
state. The two states $\Lambda(2080)5/2^-$ and $\Lambda(2100)7/2^-$  belong to the third
excitation band and are not listed in Table~\ref{Phases-Lam}. They are assumed to belong to the
SU(3) singlet series because of their devays (see below).

\begin{table*}[!h]
\caption{\label{Bonn-Sig}Configuration mixing of $\Sigma$ resonances in the Isgur-Karl (first
number) and in Bonn model $\mathcal{A}$ (second number); the fractions are given in \%, n.g. = not
given, * = our value  (assuming no mixing with the third excitation band). 
Shown are negative-parity states in the $1\hbar\omega$ band and positive-parity states in
the $2\hbar\omega$ band. The two states with a dagger $\dagger$ have two possible assignments,
they may belong to the first or to the third (unshown) excitation band.
In the $2\hbar\omega$ band, six states with $J^P=1/2^+$ are expected,
eight with $J^P=3/2^+$, five with $J^P=5/2^+$, and two with $J^P=7/2^+$. Higher-mass states are
omitted.} \footnotesize
\begin{center}
\renewcommand{\arraystretch}{1.6}
\begin{tabular}[h]{|c||c||cccc||cc||c|}
\hline
$J^\pi$&
Model state &
$\!\!{}^2 8[56]\!\!$&
$\!\!{}^2 8[70]\!\!$&
$\!\!{}^4 8[70]\!\!$&
$\!\!{}^2 8[20]\!\!$&
$\!\!{}^4 10[56]\!\!$&
$\!\!{}^2 10[70]\!\!$&Identified with\\[1mm]
\hline
\hline
     &$\MSS(1,-,1,1650/1628)$ &      n.g./5.4 & {\bf \underline{67/87.4}} &   29/2.3 &   n.g./0.1 &    n.g./0.0 &     3/3.4 & $\Sigma(1620)1/2^-$\\
\cline{2-9}
$\frac{1}{2}^-$     &$\MSS(1,-,2,1750/1771)$ &   n.g./0.2 &  21/2.9 & {\bf \underline{ 66/94.6}} &    n.g./0.2 &     n.g./0.3 &     12/1.1& $\Sigma(1750)1/2^-$\\
\cline{2-9}
     &$\MSS(1,-,3,1810/1798)$  &   n.g./0.1 &  11/2.8 &      4/1.7 &      n.g./0.3 &     n.g./0.0 & {\bf \underline{85/94.4}} &$\Sigma(1900)1/2^-\dagger$\\
\hline
\hline
     &$\MSS(3,-,1,1675/1669)$  &   n.g./5.1 & {\bf \underline{92/89.0}} &   1/1.2 &    n.g./0.1 &   n.g./0.0 &   7/3.4&$\Sigma(1670)3/2^-$ \\
\cline{2-9}
$\frac{3}{2}^-$     &$\MSS(3,-,2,1805/1728)$ &    0*/0.1 &     2*/0.1 & 41*/{\bf \underline{82.7}} &0*/0.1 &  0*/0.2 &     {\bf \underline{57*}}/16.0 & - \\
\cline{2-9}
                           &$\MSS(3,-,3,1815/1781)$  &   n.g./0.2 &      6/4.4 &     {\bf \underline{58}}/15.0 &   n.g./0.2 &     n.g./0.0 &36/{\bf \underline{79.3}} &$\Sigma(1910)3/2^-\dagger$\\
\hline
\hline
$\frac{5}{2}^-$     &$\MSS(5,-,1,1760/1770)$  &      0*/0.0 &      0*/0.0 & {\bf \underline{100*/99.0}} &   0*/0.0 &    0*/0.2 &   0*/0.0 &$\Sigma(1775)5/2^-$\\
 \hline\hline
     &$\MSS(1,+,2,1640/1760)$ & {\bf \underline{94/96.1}} &     5/2.3 &  0/0.0 &  0/0.1 &    0/0.0 &      1/0.2 & $\Sigma(1660)1/2^+$\\
 \cline{2-9}
 $\frac{1}{2}^+$  &$\MSS(1,+,3,1910/1947)$  &  4/6.9 & {\bf \underline{82/88.4}} &  9/0.9 &      0/0.3 &     1/0.0 &      3/2.5 & $\Sigma(1880)1/2^+$ \\
\cline{2-9}
   &$\MSS(1,+,4,1995/2009)$ &      1/0.0 &     7/ 0.2 &      18/8.4 &      1/0.1 & {\bf \underline{67/89.9}} &   6/0.4& \\
\cline{2-8}
     &$\MSS(1,+,5,2025/2052)$ &      1/0.8 &      0/1.8 &      15/1.2 &      2/1.9 &     0/0.2 & {\bf \underline{82/93.2}} &\\
\hline
\hline
     &$\MSS(3,+,1,1865/1896)$ & 0/{\bf \underline{73.9}} &     0/22.2 &     8/0.6 &      0/0.1 & {\bf \underline{91}}/0.0 &     0/2.0 &$\Sigma(1780)3/2^+$\\
\cline{2-9}
$\frac{3}{2}^+$      &$\MSS(3,+,2,1935/1961)$  &     {\bf \underline{86}}/0.0 &     8/0.0 &      2/5.1 &     0/0.1 & 1/{\bf \underline{93.9}} &     3/0.1 &$\Sigma(1940)3/2^+$\\
 \cline{2-9}
     &$\MSS(3,+,3,2005/2011)$  &      2/1.5 &      10/1.5 &     41/17.3 &      1/1.4 & {\bf \underline{43/73.4}} &     2/4.0 &$\Sigma(2080)3/2^+$\\
\hline
\hline
      &$\MSS(5,+,1,1940/1956)$  & {\bf \underline{88/77.8}} &     7/18.2 &      0/0.2 &     0/0.0 &     0/0.0 &     4/2.5 &$\Sigma(1915)5/2^+$\\
\cline{2-9}
$\frac{5}{2}^+$     &$\MSS(5,+,2,2035/2027)$  &     0/2.9 &     4/7.8 &     19/16.3 &      0/0.0 & {\bf \underline{77/65.9}} &      1/6.0 &{$\Sigma(2070)5/2^+$}\\
\cline{2-9}
    &$\MSS(5,+,3,2060/2071)$  &     7/14.0 & {\bf \underline{88/72.0}} &     0/7.6 &     0/0.0 &      4/4.9 &     0/0.4 &\\
\hline
\hline
\multirow{2}{*}{$\frac{7}{2}^+$}      &$\MSS(7,+,1,2015/2070)$  &    0/0.0 &      0/0.0 &     24/29.4 &      0/0.0 & {\bf \underline{76/69.6}} &     0/ 0.0 &$\Sigma(2030)7/2^+$\\
\cline{2-9}
     &$\MSS(7,+,2,2115/2161)$  &      0/0.0 &      0/0.0 & {\bf \underline{76/70.0}} &      0/0.0 &     24/29.2 &      0/0.0& \\
\hline
\end{tabular}
\vspace{2mm}
\end{center}
\end{table*}

A mismatch is $\Lambda(1710)1/2^+$. Based on the KSU analysis, it should be a SU(3) octet state,
based on Table~\ref{Bonn-Lam} the assignment to the SU(3) singlet series is preferred. It
is neither seen in any of the early analyses nor by Osaka-ANL nor by BnGa; obviously it is
difficult to extract from the data. Thus we believe it to be a SU(3) singlet state even though the
KSU analysis favors it as SU(3) octet state.

For the octet candidates, the early analyses and Kent find mostly phases that are compatible with
the octet interpretation, except for the Kent result on $\Lambda(2050)3/2^-$. For $\Lambda(1670)1/2^-$
Osaka-ANL finds, instead of the expected $180^\circ$, a value at about $-34^\circ$,
BnGa finds $\delta=(-70\pm 18)^\circ$.  This is a clear unresolved discrepancy. For
$\Lambda(1810)1/2^+$, only Osaka/ANL deviates from the expectation.

 $\Lambda(2000)1/2^-$ and $\Lambda(2050)3/2^-$ are not listed in Table~\ref{Bonn-Lam}.
About 150\,MeV below their masses,
there is a $N(1875)$ $3/2^-$. It is accompanied by $N(1895)1/2^-$ and
not by a $5/2^-$ state. Hence these two $N^*$ states form a spin-doublet. Since there is a spin-triplet
(a degenerate spin-quartet) of negative-parity $\Delta$ states close by (at 1900, 1940, 1930\,MeV
with $J^P=1/2^-, 3/2^-, 5/2^-$), we assign these three $\Delta$ and the two nucleon
resonances to a $[56]$-plet with a $^410$ quartet
and a $^28$ doublet. The $\Lambda(2000)1/2^-$ and $\Lambda(2050)3/2^-$ may form
a spin-doublet and could be the hyperon partners of the two $N^*$ states.
 Indeed, $\Lambda(2000)1/2^-$ is identified in the early analyses and by Kent
-- via its phases -- as octet state. The results on $\Lambda(2050)3/2^-$ are ambiguous.

Finally, we discuss the three states  $\Lambda(2070)3/2^+$,
$\Lambda(2110)$ $5/2^+$, and $ \Lambda(2085)7/2^+$.
At the first glance, they seem to belong to a spin-quartet $^48[70]$-plet in the second
excitation band. For $ \Lambda(2085)7/2^+$, there is no alternative interpretation. For quartet
states, no phase can be predicted on the basis of SU(6). The Bonn model suggests that the other
two states could belong to the singlet series. This is confirmed by the BnGa phases.

\subsection{\boldmath$\Lambda(1380)1/2^-$ and $\Lambda(1405)1/2^-$}
In this section, we have interpreted the $\Lambda(1405)1/2^-$ and the $\Lambda(1520)3/2^-$
resonances as $qqq$ resonances in which one of the quarks is excited to the $p$ state. However,
this interpretation is not uncontested. In modern approaches based on effective field theories, the
$\Lambda(1405)1/2^-$ emerges as quasi-bound state in the $\bar KN$ and $\pi\Sigma$ coupled-channel
system and an additional state, named the  $\Lambda(1380)1/2^-$ appears in RPP'2020.  Both states are
{\it dynamically generated}, see Refs.~\cite{Cieply:2016jby,Anisovich:2020tbd}. In a recent paper,
Mei\ss ner suggested that a hadron resonance may manifest itself as a two-pole structure~\cite{Meissner:2020khl}.

\hspace{-2mm}Decades ago, the quark model  predicted the existence of states like $N(1440)1/2^+$,
$N(1535)1/2^-$, $\Delta(1700)3/2^-$. These states have more recently also been interpreted as
dynamically generated resonances. The quark model requires one low-mass $\Lambda$ resonance with
$J^P=1/2^-$ that is dominantly an SU(3) singlet state. Models based on effective field theories
assign a large SU(3) singlet component to $\Lambda(1380)1/2^-$. In any case, one and only one of
the two  states $\Lambda(1380)1/2^-$ and $\Lambda(1405)1/2^-$ can and has to be assigned to the
predicted quark-model state. The other one - if it exists - must be an extra state, an intruder,
incompatible with any quark-model interpretation.

\begin{table}[pt]
\caption{\label{tab:S-decays}SU(3) coupling constants for $\Sigma^*$ decays and the SU(6) predictions.
\vspace{-3mm}}
\renewcommand{\arraystretch}{1.5}
\bc
\begin{tabular}{cccc}
&&&\\
\hline\hline
Decay mode                &$8\to 8+8$ & $10\to 8+8$           & \\\hline\hline
&&&\\[-3ex]
$\Sigma\to N\bar K$  &$\sqrt{2}(2\alpha -1)A_8$&$-\sqrt{\frac16}A_{10}$        & \\
$\Sigma\to \Sigma\pi$&$2\sqrt{2}\cdot\alpha A_8$             &$\sqrt{\frac16}A_{10}$& \\
$\Sigma\to \Lambda\pi$&$-\frac{2}{\sqrt{3}}(\alpha-1)A_8$             &$-\frac12 A_{10}$& \\
\hline\hline
\end{tabular}\\[2ex]
\begin{tabular}{ccccc}
&$^28[56]$& $^28[70]$ & $^48[70]$&$^{2,4}[10]$ \\\hline\hline
&&&\\[-3ex]
$\frac{A(\Sigma\to N\bar K)}{A(\Sigma\to \Sigma\pi)}$&$-\frac14$&  $\frac15$&  2 & $-1$   \\
Sign &$-$&$+$&$+$&$-$\\\hline
&&&\\[-3ex]
$\frac{A(\Sigma\to\Lambda\pi)}{A(\Sigma\to \Sigma\pi)}$&$\frac12\sqrt{\frac32}$&  $\frac15\sqrt{\frac32}$& $-\sqrt{\frac32}$  &    $-\sqrt{\frac32}$    \\
Sign &$+$&$+$&$-$&$-$\\
\hline\hline
\end{tabular}
\ec\vspace{2mm}
\end{table}

\section{\label{Conf-mix-Sig}Configuration mixing of $\Sigma$ resonances}
\subsection{\label{Comp-Sig}Comparison with quark models}
Table~\ref{Bonn-Sig} compares the experimental $\Sigma$ excitation spectrum in the first and second
excitation band with the Isgur-Karl \cite{Isgur:1978xj,Isgur:1978wd} and the Bonn
model~\cite{Loring:2001ky}. Again, not all resonances in the second band are listed. The
configuration of resonances in the third band and in higher bands is not given in the publications.

$\Sigma$ resonances with a given spin-parity $J^P$ can be mixtures of different SU(6)$\otimes$O(3)
eigenstates; they may have a symmetric or antisymmetric spatial wave function  or a wave function
with mixed symmetry. $\Sigma$ resonances can be in a flavor octet or decuplet state. For some
spin-parities, they can be in a quark-spin-doublet or quark-spin-quar\-tet. Table~\ref{Bonn-Sig}
gives the probability that a physical state with defined $J^P$ has a given set of internal quantum numbers.
In Ref.~\cite{Loring:2001ky}, the two configurations 
$^48[70]$ and $^48[56]$ are not given separately and their contributions 
given in \cite{Isgur:1978wd} are added.

In most cases, there is reasonably good agreement between the quark model calculations and the
experimental masses.  Mostly, the largest contributions, underlined in Table~\ref{Bonn-Lam}, are
the same in both calculations. However, there are a few exceptions:

In the first $\Sigma$ excitation band, we expect three  states with $J^P=1/2^-$ and $J^P=3/2^-$
and one state with $J^P=5/2^-$, see Table~\ref{Bonn-Sig}. The
states in flavor octet and with total quark spin 1/2 as dominant configuration are lower in mass; their
flavor and their spin wave functions are in a mixed symmetry configuration. One pair of quarks
is antisymmetric in spin
and flavor, this is often called a {\it good diquark}.

The other states are predicted to be close in mass. Expected are a triplet of states with
$J^P=1/2^-, 3/2^-, 5/2^-$ in the SU(3) octet, and a doublet of states $J^P=1/2^-, 3/2^-$ in the SU(3)
decuplet. These are of mixed symmetry in flavor and symmetric in the spin configuration, or
symmetric in flavor and of mixed symmetry in spin. These wave functions do not contain a good
diquark.
The mass sequence of the two upper $J^P=3/2^-$ is reversed in the two models: In the Isgur-Karl model,
the two states are very close, separated by 10\,MeV only, and seem to be strongly mixed. The higher-mass state 
has a slightly larger octet component ($^48[70]$), for the lower-mass state, the decomposition is not given in \cite{Isgur:1978xj}
but the decuplet component must prevail.  In the Bonn model, the
heavier state is the one with the larger decuplet component ($^210[70]$). Experimentally, the two higher-mass
states with $J^P=1/2^-$ and $J^P=3/2^-$ are found at a much higher mass than expected.
This might be caused by repelling forces of two
strongly-mixed states having similar masses, or there could be two or several unresolved states
which are described by one effective resonance. One $\Sigma$ resonance with $J^P=3/2^-$ 
expected in the first excitation shell, either in the $^48[70]$ or in the $^210[70]$ configuration, has not been found. 

In the second excitation band, there is again one case -- with $J^P=3/2^+$ -- where a different
level ordering is predicted in the two models. In the Bonn model, the lowest-mass state
$\Sigma(1780)3/2^+$ is in dominantly in the $^28[56]$ multiplet and interpreted as the partner
of $\Sigma(1915)5/2^+$.  The $\Sigma(1940)3/2^+$ is assigned to $^410[56]$ and would be
the state that corresponds to the Roper-like state $\Delta(1600)$ $3/2^+$. The Isgur-Karl model 
sees $\Sigma(1780)3/2^+$ as Roper-like state mainly in the $^410[56]$ multiplet and 
$\Sigma(1940)3/2^+$ in $^28[56]$ and as spin-partner of $\Sigma(1915)5/2^+$.
We think the Isgur-Karl interpretation is more likely. Radial excitations are often predicted 
at too high masses in quark models; the Roper resonance $N(1440)1/2^+$ is the best-known
example. 

The next states  $\Sigma(2080)3/2^+$ and $\Sigma(2070)5/2^+$ could be members of a 
spin-quartet with $\Sigma(2030)7/2^+$ as resonance with the largest total angular momentum,
and could belong preferentially to the $^410[56]$ multiplet. The two models are not in conflict with
this interpretation even though the mixing of the $\Sigma(2080)3/2^+$ resonance is strong in the Isgur-Karl model.
These states correspond to $\Delta(1920)3/2^+$, \ $\Delta(1905)5/2^+$, $\Delta(1950)7/2^+$ in the non-strange sector.

\subsection{\label{sec:decays}SU(3) constraints for $\Sigma^*$ decays}
Table~\ref{tab:S-decays} lists the coupling constants for $\Sigma$ resonances into $N\bar K$,
$\Sigma\pi$, and $\Lambda\pi$. The $\alpha$ ratios for the different decay modes were given in
Table~\ref{tab:L-decays}.
In Table~\ref{Phases-Sig} we compare the predicted phases with experimental values. The old analyses 
and the KSU
analysis give clear answers to the phases: they are at 0$^\circ$ or at 180$^\circ$. The Osaka-ANL
and BnGa analysis give phases at any value. Here, we give numbers which range from -180$^\circ$ to
180$^\circ$.

\begin{table*}[pt]
\caption{\label{Phases-Sig} The phase differences $\delta\phi$ between the amplitudes for
$K^-p\to\Sigma^*\to\Sigma\pi$ and $K^-p\to\Sigma^*\to \bar KN$ / $K^-p\to\Sigma^*\to\Lambda\pi$ and
$K^-p\to\Sigma^*\to\Sigma\pi$ on the real axis (2012, KSU) and at the $\Sigma^*$ pole (Osaka-ANL
and BnGa). 2012 stands for results listed in Ref.~\cite{Beringer:1900zz} when the results from
different groups are mostly consistent. The relative sign or phase is not given in
Ref.~\cite{Fernandez-Ramirez:2015tfa}. For the Osaka-ANL, the results from model $A$ and $B$  are
quoted. A index $^x$ denotes entries with a nearly vanishing amplitude. The two doublets with a
dagger $\dagger$ may have contributions from three SU(6) configurations (see Table~\ref{Sum}). See text for which phases
are to be ``expected''.}
\bc
\renewcommand{\arraystretch}{1.4}
\begin{tabular}{ccccccc}
\hline\hline
\multirow{2}{*}{$\Sigma^*$}      &\multicolumn{6}{c}{$\delta\phi$} \\
\cline{2-6}&Expected&2012&KSU&Osaka-ANL$_A$&Osaka-ANL$_B$&BnGa\\\hline
$\Sigma(1620)1/2^-$  &0/0$^\circ$         &0/0$^\circ$     &  0/0$^\circ$               &                            &-124/-45$^\circ$       &     -(133\er33)/(165\er33)$^\circ$ \\
$\Sigma(1670)3/2^-$  &0/0$^\circ$         &0/0$^\circ$      &   0/0$^\circ$              &   -57/162$^\circ$   &   14/-31$^\circ$   &     (6\er16)/-(27\er16)$^\circ$ \\\hline
$\Sigma(1750)1/2^-$  &0/180$^\circ$     &                       &       0/0$^\circ$          & -41/32$^\circ$       &                               &     -(116\er23)/(15\er23)$^\circ$ \\
$\Sigma(1775)5/2^-$  &0/180$^\circ$     & 0/180$^\circ$   &       0/180$^\circ$       & 8/-179$^\circ$        &-1/-174$^\circ$           &     (27\er16)/(120\er19)$^\circ$ \\\hline
$\Sigma(1900)1/2^-$  &$\dagger$ & 0/0$^\circ$      &     0$^\circ$/180$^\circ{}^x$ &                            &-3/-25$^\circ$           &     (5\er32)/(105\er54)$^\circ$ \\
$\Sigma(1910)3/2^-$  &$\dagger$ & 180/0$^\circ$   &                                 &                            &                               &     -(65\er62)/(-175\er29)$^\circ$ \\\hline
$\Sigma(1660)1/2^+$ &180/0$^\circ$    &                       & 180/0$^\circ{}^x$  & -90/-162$^\circ$     &  174/-50$^\circ$       &     -(45\er40)/-(150\er32)$^\circ$ \\
$\Sigma(1880)1/2^+$ &0/0$^\circ$       &  0/180$^\circ$  & 180/0$^\circ{}^x$  &         &                               &      \\
$\Sigma(1780)3/2^+$ &180/180$^\circ$ &                      &     0/0$^\circ$             &                           &                               &      \\\hline
$\Sigma(1940)3/2^+$ &180/0$^\circ$    &                       &      0/0$^\circ{}^x$&                            &                                &    \\
$\Sigma(1915)5/2^+$ &180/0$^\circ$    &180/0$^\circ$   &       180/0$^\circ$       & 172/9$^\circ$       &                               &     -(147\er19)/(10\er23)$^\circ$ \\\hline
$\Sigma(2080)3/2^+$ & 180/180$^\circ$&                      &                                 &                           &                                &                                \\
$\Sigma(2070)5/2^+$ & 180/180$^\circ$&  0$^\circ$/-     &                                 &                           &                                &     \\
$\Sigma(2030)7/2^+$ & 180/180$^\circ$&180/180$^\circ$&       180/180$^\circ$   & 173/-159$^\circ$  & 88/-87$^\circ$           &     -(157\er14)/(173\er17)$^\circ$ \\\hline
$\Sigma(2100)7/2^-$  &      &0/180$^\circ$    &       0/180$^\circ$       &                           &                                 &     (60\er25)/-(50\er32)$^\circ$ \\\hline
$\Sigma(2110)1/2^-$  & $\dagger$                      &                        &180/180$^\circ{}^x$&                           &                                &     (15\er49)/(90\er43)$^\circ$ \\
$\Sigma(2010)3/2^-$  &   $\dagger$                    &                        &                                  &                           &                                &     -(115\er33)/(40\er33)$^\circ$ \\
\hline\hline
\end{tabular}
\vspace{-4mm}
\ec
\end{table*}

\begin{table*}
\caption{\label{Sum}Comparison of the hyperon spectrum with $N$ and $\Delta$ excitations. In the
first and second excitation band, all expected states are shown. The third band lists only bands
for which at least one $\Lambda$ or $\Sigma$ candidate exists. The states with a dagger $\dagger$
are special: one pair is expected at about 1750 to 1800\,MeV, two pairs at about 2000 to 2050\,MeV.
Two pairs are found only. They are shown with two possible assignments. Likely, the two observed
pairs of states are each mixtures of these three configurations. A third pair is missing. }
\renewcommand{\arraystretch}{1.23}
\bc
\begin{tabular}{|l|c|ccc|cc|}
\hline
\multicolumn{1}{|c}{$(D,L^P_N)\ S$  $J^P$}&\multicolumn{1}{|c}{Singlet} & \multicolumn{3}{|c}{Octet}&\multicolumn{2}{|c|}{Decuplet}  \\\hline
$(56,0^+_0) \ \frac{1}{2} \ \,\frac{1}{2}^+$&& $N(939)$ &$\Lambda(1116)$ &$\Sigma(1193)$&&                     \\
\phantom{zzzzzzzz\,}$\frac{3}{2} \ \,\frac{3}{2}^+$&&               &                          &     & $\Delta(1232)$ &$\Sigma(1385)$\\\hline
$(70,1^-_1) \ \frac{1}{2}\ \, \frac{1}{2}^-$&$\Lambda(1405)$&$N(1535)$&$\Lambda(1670)$&$\Sigma(1620)$ &
$\Delta(1620)$&$\Sigma(1900)\dagger$ \\
\phantom{zzzzzzzzzzz}$\frac{3}{2}^-$&$\Lambda(1520)$ &$N(1520)$&$\Lambda(1690)$  &$\Sigma(1670)$  &$\Delta(1700)$&$\Sigma(1910)\dagger$\\
\phantom{zzzzzzzz\,}$\frac{3}{2} \ \, \frac{1}{2}^-$&&$N(1650)$&$\Lambda(1800)$&$\Sigma(1750)$   & &     \\
\phantom{zzzzzzzzzzz}$\frac{3}{2}^-$&&$N(1700)$&-&-&&\\
\phantom{zzzzzzzzzzz}$\frac{5}{2}^-$&&$N(1675)$&$\Lambda(1830)$&$\Sigma(1775)$    &&\\\hline
$(56,0^+_2) \ \frac{1}{2} \ \,\frac{1}{2}^+$&& $N(1440)$ &$\Lambda(1600)$ &$\Sigma(1660)$&                    & \\
\phantom{zzzzzzzz\,}$\frac{3}{2} \ \frac{3}{2}^+$&&               &                          &     & $\Delta(1600)$ &$\Sigma(1780)$\\\hline
$(70,0^+_2)\,\frac{1}{2} \ \,\frac{1}{2}^+$&$\Lambda(1710)$ & $N(1710)$    &$\Lambda(1810)$   &$\Sigma(1880)$ &$\Delta(1750)$&-\\
\phantom{zzzzzzzz\,}$\frac{3}{2} \ \,\frac{3}{2}^+$&& -   & -   &- &&\\\hline
$(56,2^+_2) \ \frac{1}{2} \ \frac{3}{2}^+$&&$N(1720)$  &$\Lambda(1890)$
&$\Sigma(1940)$    & &\\
\phantom{zzzzzzzz\,}$\frac{1}{2} \ \frac{5}{2}^+$&&$N(1680)$  &$\Lambda(1820)$
&$\Sigma(1915)$    & & \\
\phantom{zzzzzzzz\,}$\frac{3}{2} \ \,\frac{1}{2}^+$&& &   & &  $\Delta(1910)$   &                \\
\phantom{zzzzzzzz\,}$\frac{3}{2} \ \frac{3}{2}^+$&&  &&   & $\Delta(1920)$ &$\Sigma(2080)$\\
\phantom{zzzzzzzz\,}$\frac{3}{2} \ \frac{5}{2}^+$&&  & &   & $\Delta(1905)$ &$\Sigma(2070)$\\
\phantom{zzzzzzzz\,}$\frac{3}{2} \ \frac{7}{2}^+$&&  &  &   & $\Delta(1950)$ &$\Sigma(2030)$\\
\hline
$(70,2^+_2) \ \frac{1}{2} \ \frac{3}{2}^+$&$\Lambda(2070)$&-&-&-&-&-  \\
\phantom{zzzzzzzzzz\,}$\frac{5}{2}^+$&$\Lambda(2110)$&$N(1860)$&-&-&$\Delta(2000)$ &-  \\
\phantom{zzzzzzzz\,}$\frac{3}{2} \ \frac{1}{2}^+$&&$N(1880)$&-&-   & &  \\
\phantom{zzzzzzzzzz\,}$\frac{3}{2}^+$&&$N(1900)$&-&-&&   \\
\phantom{zzzzzzzzzz\,}$\frac{5}{2}^+$&&$N(2000)$&-&-&&   \\
\phantom{zzzzzzzzzz\,}$\frac{7}{2}^+$&&$N(1990)$&$\Lambda(2085)$&-&&   \\
\hline
$(20,1^+_2) \ \frac{1}{2} \ \frac{1}{2}^+$&-&-&-&-&&  \\
\phantom{zzzzzzzzzz\,}$\frac{3}{2}^+$&-&-&-&-& &  \\
\phantom{zzzzzzzzzz\,}$\frac{5}{2}^+$&-&&&& &  \\
\hline
$(56,1^-_3) \ \frac{1}{2} \ \frac{1}{2}^-$&&$N(1895)$&$\Lambda(2000)$   &$\Sigma(1900)\dagger$&&  \\
\phantom{zzzzzzzzzz\,}$\frac{3}{2}^-$&&$N(1875)$&$\Lambda(2050)$&$\Sigma(1910)\dagger$ &&   \\
\phantom{zzzzzzzz\,}$\frac{3}{2} \ \frac{1}{2}^-$&&&&&$\Delta(1900)$&$\Sigma(2110)\dagger$  \\
\phantom{zzzzzzzzzz\,}$\frac{3}{2}^-$&&&&&$\Delta(1940)$    &  $\Sigma(2010)\dagger$\\
\phantom{zzzzzzzzzz\,}$\frac{5}{2}^-$&&&&&$\Delta(1930)$    &- \\
\hline
$(70,3^-_3) \ \frac{1}{2} \ \frac{5}{2}^-$&$\Lambda(2080)$&$N(2060)$&-   &-&-&-   \\
\phantom{zzzzzzzzzz\,}$\frac{7}{2}^-$&$\Lambda(2100)$&$N(2190)$&-&$\Sigma(2100)$&$\Delta(2200)$
   & -  \\
\hline
\end{tabular}
\ec\vspace{-2mm}
\end{table*}
In the low-energy region, Osaka-ANL and BnGa find phases that are incompatible with the
expectation. Indeed, the $\Sigma(1620)1/2^-$ and $\Sigma(1750)1/2^-$ were difficult to separate in
the BnGa analysis, and Osaka-ANL  finds only $\Sigma(1620)1/2^-$ in model $B$ and only
$\Sigma(1750)$ $1/2^-$ in model $A$. Also $\Sigma(1660)$ $1/2^+$ does not show the expected phases,
neither in the Osaka-ANL nor in the BnGa analysis. Due to its high spin at a low mass,
$\Sigma(1775)5/2^-$ can be identified rather well in partial-wave analyses. Its SU(3) structure can
be deduced consistently from the partial wave analyses (In the BnGa analysis, the relative phases
are at least compatible with the expectations within 3$\sigma$).

In about 1800\,MeV, a further spin doublet of states with $J^P=1/2^-$ and $3/2^-$ is expected that would correspond to
$\Delta(1620)1/2^-$ and $\Delta(1700)3/2^-$. Two states are found, indeed, but at about 1900\,MeV:
$\Sigma(1900)1/2^-$ and $\Sigma(1910)3/2^-$. The conflicting results on their SU(3) structure seem
also not to be compatible with an assignment to a spin-doublet SU(3) decuplet. Likely, higher-mass
resonances and these states are not separately identified.

The $\Sigma(1915)5/2^+$ and $\Sigma(2030)7/2^+$ are both well established resonances. Their phases
are at least qualitatively compatible with their common interpretation as companion of
$N(1680)5/2^+$ and $\Delta(1950)7/2^+$.

The $\Sigma(2100)7/2^-$ resonance might be the hyperon partner of $N(2190)7/2^-$ even though
its mass would rather be expected at a 200\,MeV higher mass. Alternatively, it could belong to
the $(70, 2^-)$ multiplet. In both cases, the low mass is surprising. 

\section{Comparison of $\Lambda$ and $\Sigma$ spectrum with the $N$ and $\Delta$}

Finally, we present in Table~\ref{Sum} a comparison of the spectrum of $\Lambda$ and $\Sigma$
resonances with those for $N$ and $\Delta$ resonances. Mostly, the masses of $\Lambda$ and $\Sigma$
resonances are 100 to 200\,MeV higher than the masses of their $N$ and $\Delta$ counterparts. The
SU(3) singlet states have, of course, no states in the $N/\Delta$ sector to be compared with.

The $\Lambda$ spectrum expected in the first excitation band is nearly complete, with a $J^P=3/2^-$
$\Lambda$ as missing particle. The $\Lambda(1380)1/2^-$ and $\Lambda(1405)1/2^-$ resonances are
below the $K^-p$ threshold and were not reported (except in Ref.~\cite{Kamano:2015hxa}, model $B$
where it is seen with a mass of 1512\,MeV and with a very large width).  Based on their masses and phases, 
the other states are in most cases consistently identified as singlets or octet states.

The results on the $\Sigma$ states in the first excitation shell are much less consistent. Here,
the assignment of the four lowest-mass negative-parity states to the expected spin doublet and
triplet is plausible when the masses are considered but not on the basis of their decays. Again,
one $3/2^-$-state is missing. The states $\Sigma(1900)1/2^-$ and $\Sigma(1910)$ $3/2^-$ are
considerably more massive than their possible partners \ $\Delta(1620)1/2^-$ and
$\Delta(1700)3/2^-$. \ Likely, the $\Sigma(1900)1/2^-$ and $\Sigma(1910)3/2^-$ structures contain
more than one resonance each.

In the third excitation shell, partners of $N(1895)$ $1/2^-$, $N(1875)3/2^-$, $\Delta(1900)1/2^-$,
$\Delta(1940)3/2^-$, and $\Delta(1930)$ $5/2^-$ are expected. Their masses are well separated from
the states in the first excitation shell, $\Delta(1620)1/2^-$ and $\Delta(1700)3/2^-$, that should
be accompanied with $\Sigma$ states at about 1800\,MeV. The $\Sigma$ partners of the $N^*$ doublet
should have masses above 2000\,MeV, the partners  of the $\Delta$ triplet masses of about
2100\,MeV. We observe a doublet, $\Sigma(1900)1/2^-$ and $\Sigma(1910)3/2^-$
 instead of states at 1800\,MeV and above 2000\,MeV. When $\Sigma(1900)1/2^-$ and
$\Sigma(1910)3/2^-$ are made of more than single resonances, a clear identification of
their SU(6) structure cannot be expected. A further pair of states is seen:
$\Sigma(2110)1/2^-$ and $\Sigma(2010)3/2^-$. Thus we expect three states with $J^P=1/2^-$
and three with $J^P=3/2^-$ in the (approximate) mass range from 1800 to 2100\,MeV. Two pairs
are observed, a third pair and a $J^P=5/2^-$ state are missing. At present, it is not possible
to decide which states are seen and which ones not.

The four states $\Lambda(1600)1/2^+$, $\Lambda(1810)1/2^+$, $\Sigma(1660)$ $1/2^+$,
$\Sigma(1880)1/2^+$ are likely the SU(3) partner states of $N(1440)1/2^+$ and $N(1710)1/2^+$.
$\Sigma(1780)3/2^+$ could be the hyperon candidate for a partner of $\Delta(1600)3/2^+$.
Again, in some cases the decay modes are in conflict with these interpretations.

The spin-doublets $\Lambda(1890)3/2^+$ / $\Lambda(1820)5/2^+$ and $\Sigma(1915)5/2^+$ /
$\Sigma(1940)3/2^+$ are easily interpreted as the strange partners of $N(1720)3/2^+$ /
$N(1680)5/2^+$, this assignment follows from their masses, and is suggested in the Isgur-Karl
model. Three of these states are listed with four stars, only $\Sigma(1940)3/2^+$ is
not (yet) established. For the three established states, the phases support this assignment.
The Bonn model suggests $\Sigma(1780)3/2^+$ to be the spin partner of $\Sigma(1940)3/2^+$.
We prefer to identify  $\Sigma(1780)3/2^+$ as first decuplet radial excitation and hyperon partner
of $\Delta(1600)3/2^+$.

\hspace{-2mm}The three states $\Lambda(2070)3/2^+$,  $\Lambda(2110)5/2^+$, and $\Lambda(2085)$ $7/2^+$
are close in mass and could be interpreted as quartet of states with internal quantum numbers $L=2, S=3/2$ 
with the $J^P=1/2^+$ state missing. In the case of $\Lambda(2085)7/2^+$, this interpretation is
unambiguous. This is not the case for $\Lambda(2070)3/2^+$ and $\Lambda(2110)5/2^+$. The
Bonn model interprets these latter two states as spin doublet and
assigns them to the $^21[70]$ SU(3) singlet. The decay modes from BnGa are compatitble with this interpretation.
In the Isgur and Karl-model, the $\Lambda(2070)3/2^+$ is heavily mixed but still, the largest
contribution stems from the $^21[70]$ SU(3) singlet. For $\Lambda(2110)5/2^+$, both models
- and the decay mode analysis - agree that this is a SU(3) singlet state.  

The $\Delta(1950)7/2^+$ resonance is prominently observed in $\pi N$ induced reactions.This is the
reason why we assign $\Sigma(2030)7/2^+$ to the SU(3) decuplet, and this assignment is mostly
compatible with the edcay mode analysis. There are two states,
$\Sigma(2080)3/2^+$, $\Sigma(2070)5/2^+$, that could be the spin partners of $\Sigma(2030)7/2^+$,
with a $J^P=1/2^+$ state missing.

The states $\Lambda(2000)1/2^-$ / $\Lambda(2050)3/2^-$ and
$\Sigma(1900)$ $1/2^-$ / $\Sigma(2010)3/2^-$ may form two spin doublets and could fall, jointly with
$N(1895)1/2^-$ and $N(1875)3/2^-$ and with $\Delta(1900)1/2^-$, $\Delta(1940)3/2^-$,
$\Delta(1930)5/2^-$ into the $(56, 1^-_3)$ multiplet, with three missing $\Sigma$ states, but this
is speculative at the moment. The $(56, 1^-_3)$ multiplet houses states that carry one unit of
orbital angular momentum and one unit of radial excitation. It belongs to the third excitation
band.

In comparison to the $N$ and $\Delta$ spectrum, there are still many empty slots, and many of the
states in Table~\ref{Sum} are observed with weak evidence only. There is certainly need for new
data. Indeed, new data on hyperon spectroscopy can be expected from J-PARC~\cite{Sako:2013prop},
JLAB~\cite{Adhikari:2017prop}, and the forthcoming PANDA experiment~\cite{Iazzi:2016fzb}. Possibly,
also existing data from JLab~\cite{Moriya:2013hwg} and LHCb~\cite{Aaij:2015tga} can contribute to
hyperon spectroscopy.

\section{Summary}

The spectrum of $\Lambda$ and $\Sigma$ excitations has been re-analyzed recently by four different
groups. The analyses used different coupled-channel approaches; the resulting spectrum showed the
same leading resonances, mostly 3-star and 4-star resonances in the RPP notation, and different
sets of additional resonances. In this paper, we took into account  hyperon resonances that were
seen in one of the recent partial-wave analyses. The resulting spectrum was compared with the
Isgur-Karl model and the Bonn model. The SU(3) structure of the observed resonances was discussed
by comparison with the model calculations and by a comparison of the observed decay modes with
SU(6) phase relations. In the $\Lambda$ sector, there is reasonable agreement between the
identification of singlet or octet states based on the comparison of observed states with the quark
model and the identification based on the relative phases between decay modes. Seven $\Lambda$
states are proposed to be classified as SU(3) singlet states.  In the $\Sigma$ sector, the
identification of octet or decuplet states is consistent only for a few leading resonances. In both
sectors, the first excitation shells are filled, except for the missing $3/2^-$ states. However,
several states are seen only with poor evidence. In the second shell, there are numerous missing
states and most states are not yet established. New data are certainly utterly needed.
Nevertheless, the comparison of the resulting hyperon spectrum with the spectrum of $N$ and
$\Delta$ resonances shows evidence for SU(3) symmetry. It is remarkable that even the 1* resonances
find a slot in this comparison.

{\it
This work was supported by the \textit{Deutsche Forschungsgemeinschaft} (Bonn: SFB/TR110),
the U.S. Department of Energy grant DE-SC0016582, the JSA/
DOE Contract DE-AC05-06OR23177,
and the \textit{Russian Science Foundation} (RSF 16-12-10267) We thank J. Kohlen for
drawing the figures.
}

 \end{document}